\documentclass[aps,prd,twocolumn,groupedaddress,nofootinbib]{revtex4-1}
\usepackage{amsmath,amssymb}

\begin{document}
\title{Gravitational Wave Solutions to Linearized Jordan-Brans-Dicke Theory\\ on a Cosmological Background}
\author{Onder Dunya}
\email[]{onder.dunya@boun.edu.tr}
\author{Metin Arik}
\email[]{metin.arik@boun.edu.tr}
\affiliation{Department of Physics, Bogazici University, Bebek, Istanbul, Turkey}
\date{\today}

\begin{abstract}
Approximate vacuum solutions of Jordan-Brans-Dicke theory for perturbed scalar field and perturbed Robertson-Walker metric, are found. Solutions for the scale factor $a(t)$ and the scalar field $\phi (t)$ in unperturbed JBD theory are dependent on the $\omega$ parameter which determines how the scalar field is coupled to geometry of space-time. After adding metric perturbation $h_{\mu\nu}(x)$ to Robertson-Walker metric and perturbation $\delta\phi(x)$ to the scalar field $\phi(t)$, we solved the JBD equations such that the scale factor and the scalar field solutions are $a\propto t$ and $\phi\propto t^{-2}$ with $\omega=-3/2$. These results are necessary conditions for ordinary and scalar gravitational waves to exist in the vacuum case. Despite $\omega >10^4$ for current solar system environment observations, $\omega=-3/2$ makes JBD theory conformally invariant and fits recent supernovae type Ia data. We also looked for the value of $\omega$ for the case which has nonzero spatial curvature parameter.
\end{abstract}

\maketitle

\section{Introduction}
In Jordan-Brans-Dicke\citep{bransanddicke} theory, the gravitational constant $G$ is not a constant but a parameter and it is related to a scalar field $\phi$. So the Jordan-Brans-Dicke action contains the Lagrangian of this scalar field $\phi$ and looks like
\begin{equation}
S=\frac{1}{16\pi}\int d^4x\sqrt{-g}\left(\phi R+16\pi\mathcal{L}_M-\omega\frac{g^{\mu\nu}\partial_{\mu}\phi\partial_{\nu}\phi}{\phi}\right)
\end{equation}
where $R$ is the Ricci scalar, $\mathcal{L}_M$ is the Lagrangian of matter and $\omega$ is the JBD parameter. To get the JBD equations, we should vary the action with respect to $g^{\mu\nu}$ and $\phi$. Variation operations give us the equations as
\begin{equation}\label{2}
\begin{split}
R_{\mu\nu}-\frac{1}{2}Rg_{\mu\nu}= & \frac{8\pi}{\phi}T_{\mu\nu}+\frac{1}{\phi}\left(\nabla_\mu\partial_\nu\phi-g_{\mu\nu}g^{\alpha\beta}\nabla_\alpha\partial_\beta\phi\right)\\
& +\frac{\omega}{\phi^2}\left(\partial_\mu\phi\partial_\nu\phi-\frac{1}{2}g_{\mu\nu}g^{\alpha\beta}\partial_\alpha\phi\partial_\beta\phi\right)
\end{split}
\end{equation}
and
\begin{equation}\label{3}
R+2\frac{\omega}{\phi}g^{\mu\nu}\nabla_\mu\partial_\nu\phi-\frac{\omega}{\phi^2}g^{\mu\nu}\partial_\mu\phi\partial_\nu\phi=0
\end{equation}
where
\begin{equation}
T_{\mu\nu}=-\frac{2}{\sqrt{-g}}\frac{\delta{S_M}}{\delta{g^{\mu\nu}}}
\end{equation}
is the energy momentum tensor of matter. Contracting (\ref{2}) with $g^{\mu\nu}$ gives
\begin{equation}\label{5}
-R=\frac{8\pi}{\phi}T-\frac{3}{\phi}\nabla_\mu\partial^\mu\phi-\frac{\omega}{\phi^2}\partial_\mu\phi\partial^\mu\phi
\end{equation}
and substituting (\ref{5}) into (\ref{3}) yields
\begin{equation}\label{6}
\frac{(3+2\omega)}{\phi}g^{\mu\nu}\nabla_\mu\partial_\nu\phi=\frac{8\pi}{\phi}T\ .
\end{equation}
Equations (\ref{2}) and (\ref{6}) are the basic equations of the theory. Also the predictions of the theory are same with the predictions of Einstein field equation when $\omega\to \infty$. Current observational data for solar system environment show that $\omega > 10^4$, so the theory is indistinguishable from general relativity or deviations are so small\cite{will}.

Now, we will focus on the vacuum solutions of the equations on a cosmological background. With word "vacuum", it is meant that there is nothing in the environment which we are interested in, no matter, no radiation and no cosmological constant. Besides, of course, the universe we live in, is not a steady state universe but it is expanding with the scale factor, so we will use Robertson-Walker metric which is 
\begin{equation}
ds^2=-(dt)^2+a^2(t)[(dx^1)^2+(dx^2)^2+(dx^3)^2]
\end{equation}
where $a$ is the scale factor of the universe and a function of time. At that point space is also considered as flat which means curvature parameter $k=0$. For the vacuum case, (\ref{2}) and (\ref{6}) transform into
\begin{equation}\label{8}
\begin{split}
R_{\mu\nu}-\frac{1}{2}Rg_{\mu\nu}=&\frac{1}{\phi}\left(\nabla_\mu\partial_\nu\phi-g_{\mu\nu}g^{\alpha\beta}\nabla_\alpha\partial_\beta\phi\right)\\
&+\frac{\omega}{\phi^2}\left(\partial_\mu\phi\partial_\nu\phi-\frac{1}{2}g_{\mu\nu}g^{\alpha\beta}\partial_\alpha\phi\partial_\beta\phi\right)
\end{split}
\end{equation}
and
\begin{equation}\label{9}
g^{\mu\nu}\nabla_\mu\partial_\nu\phi=0\ .
\end{equation}
As one can see, (\ref{9}) can be substituted into (\ref{8}), and the final form of the basic JBD equation for vacuum is\\
\begin{equation}\label{10}
\begin{split}
R_{\mu\nu}-\frac{1}{2}Rg_{\mu\nu}=&\frac{1}{\phi}\nabla_\mu\partial_\nu\phi\\ &+\frac{\omega}{\phi^2}\left(\partial_\mu\phi\partial_\nu\phi-\frac{1}{2}g_{\mu\nu}g^{\alpha\beta}\partial_\alpha\phi\partial_\beta\phi\right)\ .
\end{split}
\end{equation}

All we should do, now,  is to place the Ricci tensor components and the Ricci scalar of RW metric into the equation in order to construct equations for different $\mu$ and $\nu$ values. The Ricci tensor and the Ricci scalar of the metric for $k=0$ and in cartesian coordinates, are
\begin{equation}
R_{00}=-3\frac{\ddot{a}}{a}\ ,
\end{equation}
\begin{equation}
R_{11}=R_{22}=R_{33}=(a\ddot{a}+2\dot{a}^2)\ ,
\end{equation}
\begin{equation}
R=6\left(\frac{\ddot{a}}{a}+\frac{{\dot{a}}^2}{a^2}\right)\ ,
\end{equation}
where dots represent time derivatives. We assume that the scalar field $\phi$ is only function of time, not dependent on spatial coordinates. After substituting the Ricci tensor components and the Ricci scalar into (\ref{10}), we get for $\mu=0$ and $\nu=0$
\begin{equation}\label{14} 
3\frac{\dot{a}^2}{a^2}-\frac{\partial_0^2\phi}{\phi}-\frac{\omega}{2}\frac{(\partial_0\phi)^2}{\phi^2}=0
\end{equation}
and for $\mu=1$ and $\nu=1$
\begin{equation}\label{15}
-2\frac{\ddot{a}}{a}-\frac{\dot{a}^2}{a^2}+\frac{\dot{a}}{a}\frac{\partial_0\phi}{\phi}-\frac{\omega}{2}\frac{(\partial_0\phi)^2}{\phi^2}=0\ .
\end{equation}
Other two equations for the cases $\mu=2$, $\nu=2$ and $\mu=3$, $\nu=3$ are not different from the equation of $\mu=1$, $\nu=1$, so they do not give any new information about the vacuum case solutions. Subtracting (\ref{15}) from (\ref{14}) yields
\begin{equation}\label{16}
4\frac{\dot{a}^2}{a^2}+2\frac{\ddot{a}}{a}-\frac{\partial_0^2\phi}{\phi}-\frac{\dot{a}}{a}\frac{\partial_0\phi}{\phi}=0\ .
\end{equation}

By assuming that the scalar field  and the scale factor have solutions like power of time $t$, (\ref{15}) and (\ref{16}) give us the solutions for the powers which depend on the JBD parameter $\omega$. If $\phi\propto t^s$ and $a\propto t^q$, we get
\begin{equation}\label{17}
-3q^2+2q+sq-\frac{\omega}{2}s^2=0\ ,
\end{equation}
\begin{equation}\label{18}
6q^2-2q-sq+s-s^2=0\ .
\end{equation}
Solutions to these equations for $\omega\geq-3/2$ and $\omega\neq -4/3$ are
\begin{equation}
q_{\pm}=\frac{1}{3\omega+4}\left(\omega+1\pm\sqrt{\frac{2\omega+3}{3}}\right)\ ,
\end{equation}
\begin{equation}
s_{\pm}=\frac{1\mp\sqrt{3(2\omega+3)}}{3\omega+4}\ .
\end{equation}
They also satisfy the relation
\begin{equation}
3q+s=1\ .
\end{equation}
These solutions are same with that of O'Hanlon and Tupper\citep{ohanlontupper}. Once we get the exact value of $\omega$ observationally, values of $q$ and $s$ can be determined.
\section{Vacuum Solutions to Linearized JBD Theory}
In this section, approximate solutions for the scalar field, the scale factor, ordinary gravitational wave and scalar gravitational wave, are found by regarding perturbed RW metric and perturbed scalar field. We add first order perturbations to the metric and the scalar field, and neglect all the higher order perturbations in calculations. Since RW metric is function of time, the scalar field $\phi$ is chosen to be function of time for ansatz. In addition, the first order perturbations of the metric and the scalar field are functions of time and spatial coordinates. So perturbed metric and perturbed scalar field are
\begin{equation}
g_{\mu\nu}(x)=f_{\mu\nu}(t)+h_{\mu\nu}(x)
\end{equation}
and
\begin{equation}
\Phi(x)=\phi(t)+\delta\phi(x)
\end{equation}
where
\begin{itemize}
\item$f_{\mu\nu}$ is the RW metric, 
\item$h_{\mu\nu}$ is perturbation to the metric and $|h_{\mu\nu}|\ll | f_{\mu\nu}|$,
\item$\delta\phi$ is perturbation to the scalar field and $|\delta\phi|\ll |\phi|$.
\end{itemize}
Since we are dealing with the vacuum solutions, the energy momentum tensor and its trace are equal to zero. If the equations are written with explicit form of the field $\Phi$ and the metric $g_{\mu\nu}$, they look like 
\begin{equation}\label{24}
\begin{split}
&R_{\mu\nu}-\frac{1}{2}R(f_{\mu\nu}+h_{\mu\nu})=\frac{1}{(\phi+\delta\phi)}[\nabla_\mu\partial_\nu(\phi+\delta\phi)]\\
&+\frac{\omega}{(\phi+\delta\phi)^2}[\partial_\mu(\phi+\delta\phi)\partial_\nu(\phi+\delta\phi)\\
&-\frac{1}{2}(f_{\mu\nu}+h_{\mu\nu})(f^{\alpha\beta}-h^{\alpha\beta})\partial_\alpha(\phi+\delta\phi)\partial_\beta(\phi+\delta\phi)]
\end{split}
\end{equation}
and
\begin{equation}\label{25}
\frac{1}{(\phi+\delta\phi)}(f^{\mu\nu}-h^{\mu\nu})\nabla_\mu\partial_\nu(\phi+\delta\phi)=0\ .
\end{equation}
We have used inverse of the metric in (\ref{24}) and (\ref{25}) as
\begin{equation}\label{26}
g^{\mu\nu}(x)=f^{\mu\nu}(t)-h^{\mu\nu}(x)\ .
\end{equation}

\subsection{Ricci Tensor and Ricci Scalar for Perturbed RW Metric}
The general forms\citep{weinberg}\footnote{Since there is a minus sign difference for the definition of the Riemann tensor in Weinberg's book, we multiply first order components of the Ricci tensor with a minus sign.} of first order components of the Ricci tensor are
\begin{equation}\label{27}
\delta R_{00}=-\frac{1}{2a^2}\left[\partial_0^2 h_{kk}-2\frac{\dot{a}}{a}\partial_0 h_{kk}+2\left(\frac{\dot{a}^2}{a^2}-\frac{\ddot{a}}{a}\right)h_{kk}\right]\ ,
\end{equation}
\begin{equation}\label{28}
\delta R_{0i}=-\frac{1}{2}\partial_0\left[\frac{1}{a^2}(\partial_i h_{kk}-\partial_k h_{ki})\right]\ ,
\end{equation}
\begin{equation}\label{29}
\begin{split}
\delta R_{ij}=&-\frac{1}{2a^2}\left[\nabla^2h_{ij}-\partial_j\partial_k h_{ik}-\partial_i\partial_k h_{jk}+\partial_i\partial_j h_{kk}\right]\\
&+\frac{1}{2}\partial_0^2 h_{ij}-\frac{\dot{a}}{2a}[\partial_0 h_{ij}-\delta_{ij} \partial_0 h_{kk}]\\
&+\frac{\dot{a}^2}{a^2}[2h_{ij}-\delta_{ij} h_{kk}]\ .
\end{split}
\end{equation}
Each of these components can be considered as summation of zeroth and first order parts like
\begin{equation}\label{30}
R_{\mu\nu}=\bar{R}_{\mu\nu}+\delta R_{\mu\nu}
\end{equation}
where
\begin{itemize}
\item$R_{\mu\nu}$ is the Ricci tensor for the metric $g_{\mu\nu}$,
\item$\bar{R}_{\mu\nu}$ is the Ricci tensor for RW metric,
\item$\delta R_{\mu\nu}$ is perturbation of the Ricci tensor.
\end{itemize}
Before proceeding to compute Ricci tensor components, we can make some simplifications for our sake. As is known, for Einstein field equation in Minkowski space-time, transverse-traceless perturbation $h_{\mu\nu}^{TT}$ represents plane wave solution in cartesian coordinates and it is composed of plus and cross polarized waves. For a plane wave which is propagating in $x^3$ direction, it looks like
\begin{equation}
h_{\mu\nu}^{TT}=
\begin{pmatrix}
0 & 0        & 0      & 0 \\
0 & h_{11}   & h_{12} & 0 \\
0 & h_{21}   & h_{22} & 0 \\
0 & 0        & 0      & 0
\end{pmatrix}
\end{equation}
where 
\begin{equation}
h_{11}(t-x^3)=h_{+}e^{ik_\sigma x^\sigma}\  \mathrm{and}\ h_{22}=-h_{11}
\end{equation}
and
\begin{equation}
h_{12}(t-x^3)=h_\times e^{ik_\sigma x^\sigma}\ \mathrm{and}\ h_{12}=h_{21}\ .
\end{equation}
Also, if we wanted to solve the JBD equations for perturbed Minkowski metric, we should choose our scalar field as
\begin{equation}
\Phi(x)=\phi_0+\delta\phi(x)
\end{equation}
where $\phi_0$ is constant and $\delta\phi$ is a function of time and spatial coordinates. The solution\citep{maggiorenicolis} would be
\begin{equation} 
h_{\mu\nu}=
\begin{pmatrix}
0 & 0 & 0 & 0 \\
0 & A^{(+)}-\frac{\delta\phi}{\phi_0} & A^{(\times)} & 0 \\
0 & A^{(\times)} & -A^{(+)}-\frac{\delta\phi}{\phi_0} & 0 \\
0 & 0 & 0 & 0
\end{pmatrix}
\end{equation}
where $A^{(+)}$ and $A^{(\times)}$ are ordinary gravitational waves and $\delta\phi/\phi_0$ is a scalar gravitational wave. Furthermore this metric perturbation has trace $\eta^{\mu\nu}h_{\mu\nu}=-2(\delta\phi/\phi_0)$. Taking the solutions of the JBD equations for perturbed Minkowski metric into consideration, we can assume that perturbation of RW metric for JBD theory has trace
\begin{equation}
f^{\mu\nu}h_{\mu\nu}=-2\frac{\delta\phi}{\phi}
\end{equation}
and it is transverse to propagation direction of the wave. For a wave which is propagating in $x^3$ direction, it can be regarded as
\begin{equation}
h_{\mu\nu}=a^2
\begin{pmatrix}
0 & 0 & 0 & 0 \\
0 & \left(A-\frac{\delta\phi}{\phi}\right) & B & 0 \\
0 & B & \left(-A-\frac{\delta\phi}{\phi}\right) & 0 \\
0 & 0 & 0  & 0
\end{pmatrix}
\end{equation}
where $A$, $B$ and $(\delta\phi/\phi)$ are in the wave form. As one can see, we can take $h_{0\nu}$ and $h_{3\nu}$ components of the metric perturbation as zero. This assumption will simplify our calculations and by using (\ref{27}), (\ref{28}) and (\ref{29}), components of the Ricci tensor can be written as
\begin{equation}\label{38}
\begin{split}
R_{00}=\bar{R}_{00}+\delta R_{00}=&
-\frac{3\ddot{a}}{a}-\frac{1}{2a^2}\partial_0^2(h_{11}+h_{22})\\
&+\frac{\dot{a}}{a^3}\partial_0(h_{11}+h_{22})\\
&+\left(\frac{\ddot{a}}{a^3}-\frac{\dot{a}^2}{a^4}\right)(h_{11}+h_{22})\ ,
\end{split}
\end{equation}
\begin{equation}\label{39}
\begin{split}
R_{11}=\bar{R}_{11}+\delta R_{11}=&(a\ddot{a}+2\dot{a}^2)+\frac{1}{2}\partial_0^2 h_{11}-\frac{1}{2a^2}\partial_3^2 h_{11}\\ 
&+\frac{\dot{a}}{2a}\partial_0 h_{22}+\frac{\dot{a}^2}{a^2}(h_{11}-h_{22})\ ,
\end{split}
\end{equation}
\begin{equation}\label{40}
\begin{split}
R_{22}=\bar{R}_{22}+\delta R_{22}=&(a\ddot{a}+2\dot{a}^2)+\frac{1}{2}\partial_0^2 h_{22}-\frac{1}{2a^2}\partial_3^2 h_{22}\\ 
&+\frac{\dot{a}}{2a}\partial_0 h_{11}+\frac{\dot{a}^2}{a^2}(h_{22}-h_{11})\ ,
\end{split}
\end{equation}
\begin{equation}\label{41}
\begin{split}
R_{33}=\bar{R}_{33}+\delta R_{33}=&(a\ddot{a}+2\dot{a}^2)-\frac{1}{2a^2}\partial_3^2(h_{11}+h_{22})\\
&+\frac{\dot{a}}{2a}\partial_0(h_{11}+h_{22})-\frac{\dot{a}^2}{a^2}(h_{11}+h_{22}),
\end{split}
\end{equation}
\begin{equation}\label{42}
\begin{split}
R_{03}=\bar{R}_{03}+\delta R_{03}=&-\frac{1}{2a^2}\partial_3\partial_0(h_{11}+h_{22})\\
&+\frac{\dot{a}}{a^3}\partial_3(h_{11}+h_{22})\ ,
\end{split}
\end{equation}
\begin{equation}\label{43}
\begin{split}
R_{12}=\bar{R}_{12}+\delta R_{12}=&\frac{1}{2}\partial_0^2 h_{12}-\frac{1}{2a^2}\partial_3^2 h_{12}\\
&-\frac{\dot{a}}{2a}\partial_0 h_{12}+\frac{2\dot{a}^2}{a^2} h_{12}\ ,
\end{split}
\end{equation}
\begin{equation}\label{44}
R_{01}=R_{02}=R_{13}=R_{23}=0\ .
\end{equation}

The Ricci scalar can be easily computed by contracting the Ricci tensor with the inverse of the metric and it can be regarded as summation of zeroth and first order parts like
\begin{equation}
R=\bar{R}+\delta R
\end{equation}
where
\begin{itemize}
\item$R$ is the Ricci scalar for the metric $g_{\mu\nu}$,
\item$\bar{R}$ is the Ricci scalar for Robertson-Walker metric,
\item$\delta R$ is perturbation of the Ricci scalar.
\end{itemize}
Contracting (\ref{30}) with (\ref{26}) yields
\begin{equation}
\begin{split}
R=R_{\mu\nu}g^{\mu\nu}&=(\bar{R}_{\mu\nu}+\delta R_{\mu\nu})(f^{\mu\nu}-h^{\mu\nu})\\
&=\bar{R}_{\mu\nu}f^{\mu\nu}-\bar{R}_{\mu\nu}h^{\mu\nu}+\delta R_{\mu\nu}f^{\mu\nu}\\ &=\bar{R}-\bar{R}_{\mu\nu}h^{\mu\nu}+\delta R_{\mu\nu}f^{\mu\nu}\ .
\end{split}
\end{equation}
Finally, the Ricci scalar for perturbed RW metric is found as
\begin{equation}\label{47}
\begin{split}
R=&6\left(\frac{\ddot{a}}{a}+\frac{{\dot{a}}^2}{a^2}\right)+\frac{1}{a^2}\partial_0^2(h_{11}+h_{22})\\
&-\frac{1}{a^4}\partial_3^2(h_{11}+h_{22})-2\left(\frac{\ddot{a}}{a^3}+\frac{\dot{a}^2}{a^4}\right)(h_{11}+h_{22})\ .
\end{split}
\end{equation}
\subsection{Solutions to Linearized JBD Equations}
In this section, our plan is to construct and solve perturbed JBD equations. Since we have found necessary elements in previous sections, they can be now placed into the equations. Then solutions of $a$, $\phi$ and the perturbations can be obtained. We have two basic equations, however the first one which is equation (\ref{24}), will yield more than one due to different components of the Einstein tensor which is $G_{\mu\nu}=R_{\mu\nu}-\frac{1}{2}R g_{\mu\nu}$. Let us begin with the second JBD equation which is (\ref{25}). There is summation in this relation between metric components and derivatives of $\Phi$, and it can be expanded as
\begin{equation}\label{48}
\begin{split}
\frac{1}{(\phi+\delta\phi)}[&f^{00}\nabla_0\partial_0(\phi+\delta\phi)+(f^{11}-h^{11})\nabla_1\partial_1(\phi+\delta\phi)\\
&+(f^{22}-h^{22})\nabla_2\partial_2(\phi+\delta\phi)\\
&+f^{33}\nabla_3\partial_3(\phi+\delta\phi)]=0\ .
\end{split}
\end{equation}
Then inserting metric components and covariant derivatives of partial derivatives of $\Phi$ into (\ref{48}) gives
\begin{equation}\label{49}
\begin{split}
&-\frac{\partial_0^2\phi}{(\phi+\delta\phi)}-\frac{3\dot{a}}{a}\frac{\partial_0\phi}{(\phi+\delta\phi)}-\frac{1}{2a^2}\frac{\partial_0\phi}{\phi}\partial_0(h_{11}+h_{22})\\
&+\frac{\dot{a}}{a^3}\frac{\partial_0\phi}{\phi}(h_{11}+h_{22})-\frac{\partial_0^2\delta\phi}{\phi}-\frac{3\dot{a}}{a}\frac{\partial_0\delta\phi}{\phi}+\frac{1}{a^2}\frac{\partial_3^2\delta\phi}{\phi}=0\ .
\end{split}
\end{equation}
To separate zeroth and first order terms, we need one more arrangement like 
\begin{equation}\label{50}
(\phi+\delta\phi)^{-1}\simeq\frac{1}{\phi}\left(1-\frac{\delta\phi}{\phi}\right)\ .
\end{equation}
After placing (\ref{50}) into (\ref{49}), the first order equation is
\begin{equation}\label{51}
\begin{split}
&-\frac{\partial_0^2\delta\phi}{\phi}-\frac{3\dot{a}}{a}\frac{\partial_0\delta\phi}{\phi}+\frac{1}{a^2}\frac{\partial_3^2\delta\phi}{\phi}\\
&-\frac{1}{2a^2}\frac{\partial_0\phi}{\phi}\partial_0(h_{11}+h_{22})+\frac{\dot{a}}{a^3}\frac{\partial_0\phi}{\phi}(h_{11}+h_{22})\\
&+\frac{\partial_0^2\phi\delta\phi}{\phi^2}+\frac{3\dot{a}}{a}\frac{\partial_0\phi\delta\phi}{\phi^2}=0\ .
\end{split}
\end{equation}
Since we know $h_{11}+h_{22}=-2a^2(\delta\phi/\phi)$, by using this relation and integration by parts method, (\ref{51}) can be written as
\begin{equation}\label{52}
\begin{split}
&-\partial_0^2\left(\frac{\delta\phi}{\phi}\right)-\frac{3\dot{a}}{a}\partial_0\left(\frac{\delta\phi}{\phi}\right)+\frac{1}{a^2}\partial_3^2\left(\frac{\delta\phi}{\phi}\right)\\
&-\frac{\partial_0\phi}{\phi}\partial_0\left(\frac{\delta\phi}{\phi}\right)=0\ .
\end{split}
\end{equation} 
This is the final form of the first order part of (\ref{48}) and it has the form of a wave equation for $(\delta\phi/\phi)$. While first three terms come from $f^{\mu\nu}\nabla_\mu\nabla_\nu(\delta\phi/\phi)$, the fourth one is an extra term.

Now, there is one more equation to solve for perturbed JBD theory. In total, this one gives six equations because there are six different nonzero components of the Einstein tensor $G_{\mu\nu}$ for perturbed RW metric. We start with inserting (\ref{38}) and (\ref{47}) into (\ref{24}) to get equation of $\mu=0$, $\nu=0$ as
\begin{equation}
\begin{split}
&3\dot{a}^2-\frac{1}{2a^2}\partial_3^2(h_{11}+h_{22})+\frac{\dot{a}}{a}\partial_0(h_{11}+h_{22})\\
&-\frac{2\dot{a}^2}{a^2}(h_{11}+h_{22})=a^2\frac{\partial_0^2\phi}{\phi}-a^2\frac{\partial_0^2\phi\delta\phi}{\phi^2}+a^2\frac{\partial_0^2\delta\phi}{\phi}\\
&+\omega a^2\left[\frac{(\partial_0\phi)^2}{2\phi^2}-\frac{(\partial_0\phi)^2\delta\phi}{\phi^3}+\frac{\partial_0\phi\partial_0\delta\phi}{\phi^2}\right]\ .
\end{split}
\end{equation}
The first order part of this equation can be written as
\begin{equation}\label{54}
\begin{split}
&-a^2\frac{\partial_0^2\delta\phi}{\phi}-\frac{1}{2a^2}\partial_3^2(h_{11}+h_{22})+\frac{\dot{a}}{a}\partial_0(h_{11}+h_{22})\\
&-\frac{2\dot{a}^2}{a^2}(h_{11}+h_{22})=-a^2\frac{\partial_0^2\phi\delta\phi}{\phi^2}\\
&+\omega a^2\left[-\frac{(\partial_0\phi)^2\delta\phi}{\phi^3}+\frac{\partial_0\phi\partial_0\delta\phi}{\phi^2}\right]\ .
\end{split}
\end{equation}
By using $h_{11}+h_{22}=-2a^2(\delta\phi/\phi)$, (\ref{52}) and the relations via integration by parts
\begin{equation}
\frac{\partial_0^2\delta\phi}{\phi}=\partial_0^2\left(\frac{\delta\phi}{\phi}\right)+2\frac{\partial_0\phi}{\phi}\partial_0\left(\frac{\delta\phi}{\phi}\right)+\frac{\partial_0^2\phi\delta\phi}{\phi^2}\ ,
\end{equation} 
\begin{equation}
\omega a^2\left[-\frac{(\partial_0\phi)^2\delta\phi}{\phi^3}+\frac{\partial_0\phi\partial_0\delta\phi}{\phi^2}\right]=\omega a^2\frac{\partial_0\phi}{\phi}\partial_0\left(\frac{\delta\phi}{\phi}\right)\ ,
\end{equation}
equation (\ref{54}) can be simplified firstly as
\begin{equation}
a\dot{a}\partial_0\left(\frac{\delta\phi}{\phi}\right)-a^2\frac{\partial_0\phi}{\phi}\partial_0\left(\frac{\delta\phi}{\phi}\right)=\omega a^2\frac{\partial_0\phi}{\phi}\partial_0\left(\frac{\delta\phi}{\phi}\right)\ ,
\end{equation}
then as
\begin{equation}\label{58}
\frac{\dot{a}}{a}-\frac{\partial_0\phi}{\phi}=\omega\frac{\partial_0\phi}{\phi}\ .
\end{equation}
We have assumed that $\phi$ and $a$ have power-law solutions like $\phi\propto t^s$ and $a\propto t^q$. So, after we put them into (\ref{58}), it turns into
\begin{equation}\label{59}
q=s(\omega+1)\ .
\end{equation}
This is a new relation of $q$, $s$ and $\omega$. When equations for different $\mu$ and $\nu$ cases are checked, this relation consistently appears . Now, values of $q$, $s$ and $\omega$ can be found by using the relations 
\begin{equation*}
-3q^2+2q+sq-\frac{\omega}{2}s^2=0\ ,
\end{equation*}
\begin{equation*}
6q^2-2q-sq+s-s^2=0\ ,
\end{equation*}
\begin{equation*}
q=s(\omega+1)\ ,
\end{equation*}
which are (\ref{17}), (\ref{18}) and (\ref{59}) respectively. The solutions that satisfy these relations are $q=1$, $s=-2$ with $\omega=-3/2$. Thus, we can write
\begin{equation}\label{60}
a(t)=a_0\left(\frac{t}{t_0}\right)
\end{equation}
and
\begin{equation}\label{61}
\phi(t)=\phi_0\left(\frac{t}{t_0}\right)^{-2}\ .
\end{equation}

Our finding for $\omega$ may seem unpleasant because it is a negative coupling parameter and solar system observations showed $\omega > 10^4$. However, JBD theory with negative $\omega$ value can explain accelerating expansion of the universe without any necessity of cosmological constant\cite{bertolami}\cite{senseshadri}\cite{banerjeepavon}. In addition, $\omega=-3/2$ is the value which makes JBD theory conformally invariant\cite{dabrowski} and fits recent data of type Ia supernovae\cite{fabris}.

Since we obtained power law solutions for the scale factor and the scalar field, it is easy to put them into the wave equation and look for the solution. Wave equation form of a scalar gravitational wave in (\ref{52}) is exactly the same for ordinary gravitational waves $A$ and $B$. Inserting (\ref{60}) and (\ref{61}) into the wave equation of $A$ yields
\begin{equation}
\frac{1}{a^2}\frac{\partial^2A}{\partial z^2}-\frac{\partial^2 A}{\partial t^2}-\frac{1}{t}\frac{\partial A}{\partial t}=0\ .
\end{equation}
This equation can be arranged as
\begin{equation}
\frac{\partial^2 A}{\partial z^2}=\left(\frac{a_0}{t_0}\right)^2 t^2 \frac{\partial^2 A}{\partial t^2}+\left(\frac{a_0}{t_0}\right)^2 t\frac{\partial A}{\partial t}\ .
\end{equation}
Simplifying the form of the wave equation by defining conformal time $\tau$, will help us to figure it out. We define
\begin{equation}
\frac{\partial t}{t}=\partial \tau
\end{equation}
so
\begin{equation}
\ln t=\tau\ .
\end{equation}
The wave equation with respect to $\tau$ is
\begin{equation}\label{66}
\frac{\partial^2 A}{\partial z^2}=\left(\frac{a_0}{t_0}\right)^2 \frac{\partial^2 A}{\partial \tau^2}+\left(\frac{a_0}{t_0}\right)^2 \frac{\partial A}{\partial \tau}\ .
\end{equation}
At that point, to be able to guess the form of the wave equation, the distance relation on a null geodesic for the scale factor can be written as 
\begin{equation}
d=\int\frac{\partial t}{a}=\frac{t_0}{a_0}\int\frac{\partial t}{t}=t_0\ln t=t_0\tau
\end{equation}
where $a_0=1$. This and the form of the wave equation in (\ref{66}) encourage us to write a wave function like
\begin{equation}
A(z,\tau)=A_0 e^{i(\tilde{k}az-kt_0\tau)}
\end{equation}
where $\tilde{k}$ is a complex wave number. Substituting this into (\ref{66}) gives
\begin{equation}
\tilde{k}^2=\left(\frac{t_0}{t}\right)^2 k^2+i\left(\frac{t_0}{t}\right)^2 \frac{k}{t_0}
\end{equation}
and so
\begin{equation}
\tilde{k}=\frac{t_0}{t}k\left(1+\frac{i}{kt_0}\right)^{1/2}=\frac{t_0}{t}k+\frac{i}{2t}\ .
\end{equation}
After inserting this back to the wave function, it becomes
\begin{equation}
A(z,\tau)=A_0 e^{i(kz-kt_0\tau)}e^{-\frac{z}{2t_0}}\ .
\end{equation}
This is the form of the wave function for ordinary and scalar gravitational waves that we are looking for. As is expected, it decays exponentially at large distances.

\section{Vacuum Solutions to JBD Theory for Non-Flat Spacelike Sections}
So far, we dealt with JBD theory on flat space and found JBD parameter $\omega=-3/2$. Now, we wonder the value of $\omega$ for cases which have nonzero curvature parameters. To find it, a new action which contains a potential term of the scalar field is the starting point. Potential term is necessary in non-flat cases, otherwise equations give the curvature parameter as zero. So, the JBD action with a self-interacting potential for the vacuum case is 
\begin{equation}
S=\frac{1}{16\pi}\int d^4x\sqrt{-g}\left(\phi R-\omega\frac{g^{\mu\nu}\partial_{\mu}\phi\partial_{\nu}\phi}{\phi}-\lambda\phi^2\right)\ .
\end{equation}
Taking variations and applying the same procedure which we did before, give the JBD equations as
\begin{equation}\label{73}
\begin{split}
R_{\mu\nu}-\frac{1}{2}Rg_{\mu\nu}= &\frac{1}{\phi}\left(\nabla_\mu\partial_\nu\phi-g_{\mu\nu}g^{\alpha\beta}\nabla_\alpha\partial_\beta\phi\right)\\
& +\frac{\omega}{\phi^2}\left(\partial_\mu\phi\partial_\nu\phi-\frac{1}{2}g_{\mu\nu}g^{\alpha\beta}\partial_\alpha\phi\partial_\beta\phi\right)\\
&-\frac{g_{\mu\nu}}{2}\lambda\phi\ ,
\end{split}
\end{equation}
\begin{equation}\label{74}
\frac{(3+2\omega)}{\phi}g^{\mu\nu}\nabla_\mu\partial_\nu\phi=0\ .
\end{equation}
We will use the following form of RW metric in spherical coordinates
\begin{equation}
ds^2=-dt^2+a^2\left(\frac{dr^2}{1-\kappa r^2}+r^2(d\theta^2+\sin^2\theta d\phi^2)\right)
\end{equation}
where $\kappa$ is curvature. Ricci tensor components and the Ricci scalar for this metric form are
\begin{equation}\label{76}
R_{00}=-3\frac{\dot{a}}{a}\ ,
\end{equation}
\begin{equation}\label{77}
R_{11}=\frac{a\ddot{a}+2\dot{a}^2+2\kappa}{1-\kappa r^2}\ ,
\end{equation}
\begin{equation}\label{78}
R_{22}=r^2(a\ddot{a}+2\dot{a}^2+2\kappa)\ ,
\end{equation}
\begin{equation}\label{79}
R_{33}=r^2\sin^2\theta(a\ddot{a}+2\dot{a}^2+2\kappa)\ ,
\end{equation}
\begin{equation}\label{80}
R=6\left(\frac{\ddot{a}}{a}+\frac{\dot{a}^2}{a^2}+\frac{\kappa}{a^2}\right)\ .
\end{equation}
After taking the derivatives, (\ref{74}) becomes
\begin{equation}
\partial_0^2\phi+3\frac{\dot{a}}{a}\partial_0\phi=0\ . 
\end{equation}
Using $a\propto t^q$ and $\phi\propto t^s$ in this equation gives $3q+s=1$ which we are familiar with from the preceding sections. Let us continue with equation (\ref{73}). By using (\ref{74}), (\ref{76}) and (\ref{80}), it can be written for $\mu=0$, $\nu=0$ as
\begin{equation}\label{82}
3\frac{\dot{a}^2}{a^2}+3\frac{\kappa}{a^2}-\frac{\partial_0^2\phi}{\phi}-\frac{\omega}{2}\frac{(\partial_0\phi)^2}{\phi^2}-\frac{\lambda\phi}{2}=0\ .
\end{equation}
For $\mu=1$, $\nu=1$ case, (\ref{73}) turns into
\begin{equation}\label{83}
-2\frac{\ddot{a}}{a}-\frac{\dot{a}^2}{a^2}-\frac{\kappa}{a^2}+\frac{\dot{a}}{a}\frac{\partial_0\phi}{\phi}-\frac{\omega}{2}\frac{(\partial_0\phi)^2}{\phi^2}+\frac{\lambda\phi}{2}=0\ .
\end{equation}
To get two independent equations, (\ref{83}) is subtracted from (\ref{82}), then we have
\begin{equation}\label{84}
2\frac{\ddot{a}}{a}+4\frac{\dot{a}^2}{a^2}+4\frac{\kappa}{a^2}-\frac{\partial_0^2\phi}{\phi}-\frac{\dot{a}}{a}\frac{\partial_0\phi}{\phi}-\lambda\phi=0\ .
\end{equation}
Since $a\propto t^q$ and $\phi\propto t^s$, after they are inserted into (\ref{83}) and (\ref{84}), following two are obtained
\begin{equation}\label{85}
\frac{1}{t^2}(6q^2-2q-sq+s-s^2)+4\frac{\kappa}{a^2}-\lambda\phi=0\ ,
\end{equation}
\begin{equation}\label{86}
\frac{1}{t^2}\left(3q^2-2q-sq+\frac{\omega}{2}s^2\right)+\frac{\kappa}{a^2}-\frac{\lambda\phi}{2}=0\ .
\end{equation}
Terms in parenthesis evolve with $1/t^2$ and it is known that the equations should be satisfied in any time. So in this case, the scale factor and the scalar field should be $a\propto t$ and $\phi\propto t^{-2}$ respectively as we found before. After placing $q=1$ and $s=-2$ into (\ref{85}) and (\ref{86}), the equations, which show the relations between parameters, are  
\begin{equation}
4\frac{\kappa}{a^2}=\lambda\phi
\end{equation}
and
\begin{equation}
\frac{3+2\omega}{t^2}+\frac{\kappa}{a^2}=\frac{\lambda\phi}{2}\ .
\end{equation}
Finally, by using the forms of the scale factor and the scalar field in (\ref{60}) and (\ref{61}), $\omega$ can be written as
\begin{equation}\label{89}
\omega=\frac{\kappa t_0^2}{2}-\frac{3}{2}\ .
\end{equation}

According to this relation, $\omega$ is still a negative coupling parameter since the size of the universe is greater than the age of the universe. Besides, obtaining true value of $\omega$ provides the ratio between them. As it is indicated in \cite{fabris}, based on the recent observational data, the best fitting value of $\omega$ is $-1.477$. So, ({\ref{89}) is a very reasonable relation. As we have referred before, although local tests indicate that $\omega$ is a large positive number, JBD theory with a negative coupling parameter is viable scenario.  
\nocite{*}
\bibliography{references}

\end{document}